\documentclass[12pt]{article}
\usepackage[dvips]{graphicx}
\usepackage{epsfig}
\usepackage[left]{lineno}

\begin{document}

\begin{center}
EUROPEAN ORGANIZATION FOR NUCLEAR RESEARCH\\
\end{center}
\begin{flushright}
CERN-PH-EP/2007-002\\
13 February 2007
\end{flushright}

%\begin{flushright}
%{\bf\underline{DRAFT 5: 28/03/2007}}
%\end{flushright}

\begin{center}
{\Large\bf Measurement of the Dalitz plot slope parameters of the
$K^\pm\to\pi^\pm\pi^+\pi^-$ decay}
\end{center}
%{\bf Corresponding author:} Dr Evgueni Goudzovski\\
%INFN, Largo B. Pontecorvo, 3, Pisa 56127 Italy\\
%tel.: +39 334 3190280, fax: +39 050 2214333\\
%e-mail: goudzovs@mail.cern.ch\\
%\\
%{\bf Key words:} Kaon hadronic decay, Dalitz plot slope, CERN SPS.\\
\begin{center}
{\Large The NA48/2 Collaboration}\\
\vspace{2mm}
 J.R.~Batley,
 A.J.~Culling,
 G.~Kalmus,
 C.~Lazzeroni,
 D.J.~Munday,
 M.W.~Slater,
 S.A.~Wotton \\
{\em \small Cavendish Laboratory, University of Cambridge,
Cambridge, CB3 0HE,
U.K.$\,$\footnotemark[1]} \\[0.2cm]
 R.~Arcidiacono,
 G.~Bocquet,
 N.~Cabibbo,
 A.~Ceccucci,
 D.~Cundy$\,$\footnotemark[2],
 V.~Falaleev,
 M.~Fidecaro,
 L.~Gatignon,
 A.~Gonidec,
 W.~Kubischta,
 A.~Norton,
 A.~Maier,
 M.~Patel,
 A.~Peters\\
{\em \small CERN, CH-1211 Gen\`eve 23, Switzerland}
% ...
\\[0.5cm]
\vspace{7mm} \it{Submitted to Physics Letters B.} \rm
\end{center}

\setcounter{footnote}{0}
\footnotetext[1]{Funded by the U.K.
Particle Physics and Astronomy Research Council}
\footnotetext[2]{Present address: Istituto di Cosmogeofisica del CNR
di Torino, I-10133 Torino, Italy}
\footnotetext[3]{Present address: Scuola Normale Superiore,
I-56100 Pisa, Italy}
\footnotetext[4]{Present address: CERN, CH-1211 Gen\`eve 23, Switzerland}
\footnotetext[5]{Also at Centre de Physique des Particules de Marseille,
IN2P3-CNRS, Universit\'e de la M\'editerran\'ee, Marseille, France}
\footnotetext[6] {Also Istituto di Fisica, Universit\`a di Modena,
I-41100 Modena, Italy}
\footnotetext[7]{Istituto di Fisica,
Universit\`a di Urbino, I-61029  Urbino, Italy}
\footnotetext[8]{Funded by the German Federal Minister for Education
and research under contract 05HK1UM1/1}
\footnotetext[9]{Permanent address: SLAC, Stanford University, Menlo
Park, CA 94025, U.S.A.}
\footnotetext[10]{Present address: UCLA,  Los Angeles, CA 90024,
U.S.A.}%
\footnotetext[11]{Funded by the German Federal Minister for
Research and Technology (BMBF) under contract 056SI74}
\footnotetext[12]{Funded by the Austrian Ministry for Traffic and
Research under the contract GZ 616.360/2-IV GZ 616.363/2-VIII, and
by the Fonds f\"ur Wissenschaft und Forschung FWF Nr.~P08929-PHY}

%%%%%%%%%%%%%%%%%%%%%%%%%%%%%%%%%%%%%%%%%%%%%%%%%%%%%%%%%%%%%%%%%
%\begin{linenumbers}

\begin{abstract}
The distribution of the $K^\pm\to\pi^\pm\pi^+\pi^-$ decays in the
Dalitz plot has been measured by the NA48/2 experiment at the CERN
SPS with a sample of $4.71\times 10^8$ fully reconstructed events.
With the standard Particle Data Group parameterization the following
values of the slope parameters were obtained:
$g=(-21.134\pm0.017)\%$, $h=(1.848\pm0.040)\%$,
$k=(-0.463\pm0.014)\%$. The quality and statistical accuracy of the
data have allowed an improvement in precision by more than an order
of magnitude, and are such as to warrant a more elaborate
theoretical treatment, including pion-pion rescattering, which is in
preparation.
\end{abstract}

%%%%%%%%%%%%%%%%%%%%%%%%%%%%%%%%%%%%%%%%%%%%%%%%%%%%%%%%%%%%%%%%%

\newpage
\section*{Introduction}

The $K^\pm\to\pi^\pm\pi^+\pi^-$ decay can be described~\cite{pdg} in
terms of the Lorentz invariant kinematic variables $u$ and $v$
defined as
\begin{equation}
u=\frac{s_{12}-s_0}{m_\pi^2},~~~~v=\frac{s_{13}-s_{23}}{m_\pi^2},
\end{equation}
$$
s_{ij}=(P_i+P_j)^2,~~~i,j=1,2,3,~~~i<j;~~~
s_0=\frac{1}{3}(s_{12}+s_{13}+s_{23})\equiv\frac{1}{3}m_K^2+m_\pi^2.
$$
Here $m_\pi$ is the charged pion mass, $m_K$ is the charged kaon
mass, $P_i$ are the pion four-momenta, the indices $i,j=1,2$
correspond to the two pions of the same electric charge (``even''
pions), and the index $i,j=3$ to the other (``odd'') pion. The
experimental Dalitz plot distribution of the
$K^\pm\to\pi^\pm\pi^+\pi^-$ decay has been up to now experimentally
analyzed~\cite{fo72,de77} in terms of a polynomial expansion in
powers of $u$ and $v$:
\begin{equation}
|M(u,v)|^2\sim C(u,v)\cdot(1+gu+hu^2+kv^2).
\label{slopes}
\end{equation}
Here $g$, $h$, $k$ are the linear and quadratic slope parameters
(terms proportional to odd powers of $v$ are forbidden by Bose
symmetry, and the variable $v$ is defined only up to a sign), and
$C(u,v)$ is the Coulomb factor~\cite{be93}:
\begin{equation}
C(u,v)=\prod_{i,j=1,2,3;~i<j} \{n_{ij}/(e^{n_{ij}}-1)\},~~~
n_{ij}=2\pi\alpha q_iq_j/\beta_{ij}, \label{coulomb}
\end{equation}
where $q_i=\pm1$ are the pion charges, $\alpha$ is the fine
structure constant, and $\beta_{ij}$ is the relative velocity of the
pions $i$ and $j$, expressed via the squared invariant mass of the
pion pair $s_{ij}$ as
\begin{equation}
\beta_{ij}=\left(1-\frac{4m_\pi^2}{s_{ij}}\right)^{1/2}
\left(1-\frac{2m_\pi^2}{s_{ij}}\right)^{-1}.
\end{equation}

Among the measurements of the slope parameters performed in the
past, the most precise are reported in~\cite{fo72} based on
$1.5\times 10^6$ $K^\pm$ decays, and in~\cite{de77} based on
$0.225\times10^6$ $K^+$ decays.

The primary goal of the NA48/2 experiment at CERN SPS is the search
for direct CP-violating charge asymmetries of Dalitz plot linear
slopes in $K^\pm\to\pi^\pm\pi^+\pi^-$~\cite{k3pic} and
$K^\pm\to\pi^\pm\pi^0\pi^0$~\cite{k3pin} decays. However, the large
data sample collected also allows a study of the Dalitz plot
distributions to be performed at a new level of precision,
estimating the detector acceptance with a detailed Monte Carlo (MC)
simulation.

The parameterization (\ref{slopes}) of the
$K^\pm\to\pi^\pm\pi^+\pi^-$ decay distribution takes into account
electromagnetic interactions only in the first
approximation~\cite{be93} and totally neglects pion rescattering
effects~\cite{ca05}--\cite{ge07}, which were recently shown by
NA48/2 to contribute significantly to the distribution of
$K^\pm\to\pi^\pm\pi^0\pi^0$ decays~\cite{cusp}. The current study
aims to measure the Dalitz plot slopes within the conventional
framework (\ref{slopes}), and represents a first step towards a more
complete description of the decay distribution. In particular, the
presence of radiative effects and strong final state interactions,
both ignored in the current work, imply that the slope parameters
obtained, while fully comparable to those defined in the Particle
Data Group compilation~\cite{pdg}, should not be attributed any
precise physical significance as far as the parameterization of the
weak decay in terms of more fundamental parameters is concerned;
this would require an improved theoretical framework which is just
being developed~\cite{co06}, and its implementation is postponed for
a forthcoming analysis.

NA48/2 collected data during two runs in 2003 and 2004, with about
50 days of efficient data-taking in each run. This analysis is based
on $4.71\times 10^8$ fully reconstructed $K^\pm\to\pi^\pm\pi^+\pi^-$
events, corresponding to 55\% of the data sample collected during
the NA48/2 run of 2003.

\section{Beams and detectors}

\begin{figure}[tb]
\vspace{-6mm}
\begin{center}
{\resizebox*{\textwidth}{!}{\includegraphics{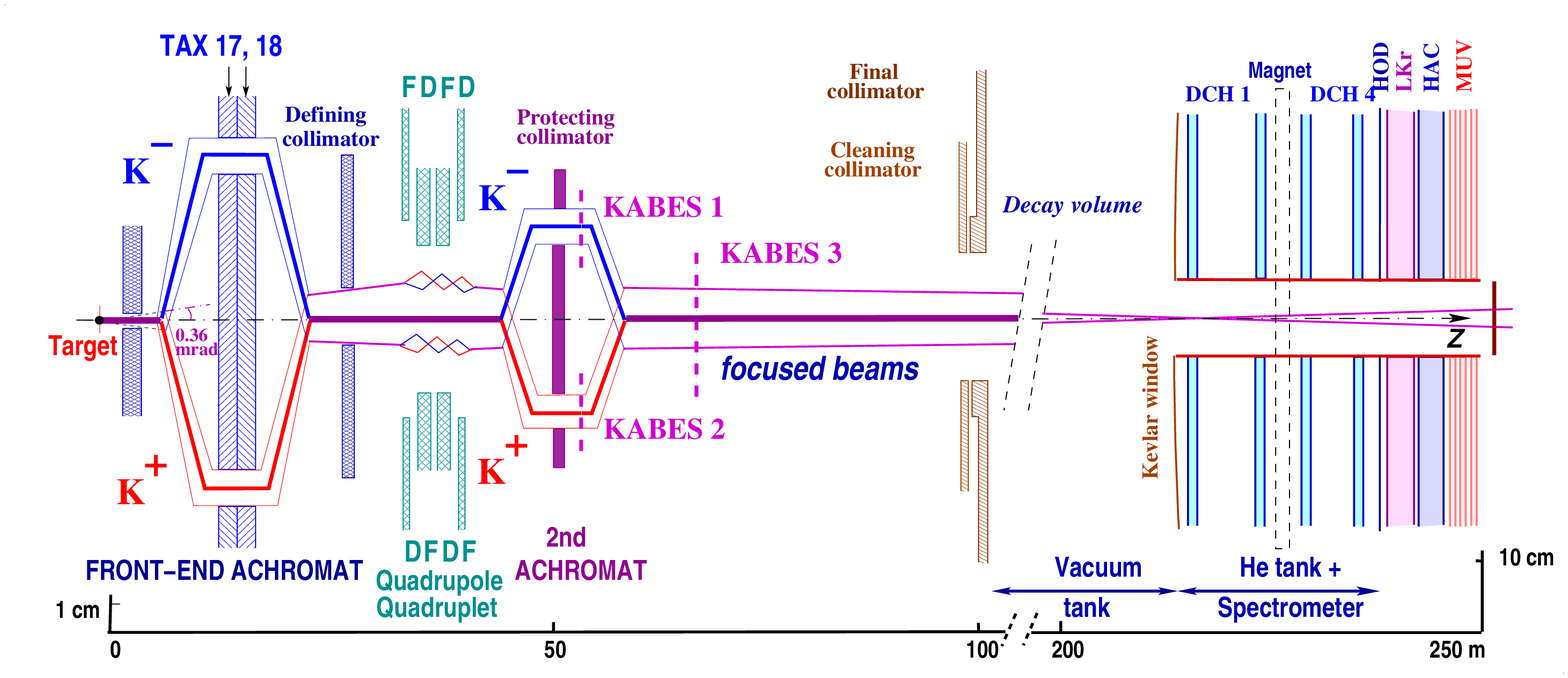}}}
\end{center}
\vspace{-6mm} \caption{Schematic lateral view of the NA48/2 beam
line (TAX17,18: motorized beam dump/collimators used to select the
momentum of the $K^+$ and $K^-$ beams; FDFD/DFDF: focusing set of
quadrupoles, KABES1--3: kaon beam spectrometer stations), decay
volume and detector (DCH1--4: drift chambers, HOD: hodoscope, LKr:
EM calorimeter, HAC: hadron calorimeter, MUV: muon veto). Note that
the vertical scales are different in the two parts of the figure.}
\label{fig:beams}
\end{figure}

A beam line providing two simultaneous charged beams of opposite
signs overlapping in space all along the decay volume was designed
and built in the high intensity hall (ECN3) at the CERN SPS. The
beam line is a key element of the experiment, as it allows decays of
$K^+$ and $K^-$ to be recorded at the same time, and therefore leads
to cancellation of several systematic uncertainties for the charge
asymmetry measurement. Regular alternation of magnetic fields in all
the beam line elements was adopted. The layout of the beams and
detectors is shown schematically in Fig.~\ref{fig:beams}. The setup
is described in a right-handed orthogonal coordinate system with the
$z$ axis directed downstream along the beam, and the $y$ axis
directed vertically up.

The beams are produced by 400 GeV/$c$ protons impinging at zero
incidence angle on a beryllium target of 40 cm length and 2 mm
diameter. Charged particles with momentum $(60\pm3)$ GeV/$c$ are
selected in a charge-symmetric way by an achromatic system of four
dipole magnets with zero total deflection (`achromat'), which splits
the two beams in the vertical plane and then recombines them on a
common axis. Then the beams pass through a defining collimator and a
series of four quadrupoles designed to produce focusing of the beams
towards the detector. Finally the two beams are again split in the
vertical plane and recombined in a second achromat, where three
stations of a Micromega-type~\cite{gi96} detector form a kaon beam
spectrometer~\cite{pe04} (not used in the present analysis).

After passing the cleaning and the final collimators, the beams
enter the decay volume housed in a 114 m long cylindrical vacuum
tank with a diameter of 1.92 m for the first 65 m, and 2.4 m for the
rest. Both beams follow the same path in the decay volume: their
axes coincide within 1~mm, while the transverse size of the beams is
about 1~cm. With $7\times 10^{11}$ protons incident on the target
per burst of $\sim 4.8$ s duration, the positive (negative) beam
flux at the entrance of the decay volume is $3.8\times 10^7$
($2.6\times 10^7$) particles per pulse, of which $5.7\%$ ($4.9\%$)
are $K^+$ ($K^-$). The $K^+/K^-$ flux ratio is about $1.8$. The
fraction of beam kaons decaying in the decay volume at nominal
momentum is $22\%$.

The decay volume is followed by a magnetic spectrometer used for
reconstructing $K^\pm\to\pi^\pm\pi^+\pi^-$ decays. The spectrometer
is housed in a tank filled with helium at nearly atmospheric
pressure, separated from the vacuum tank by a thin ($0.31\%X_0$)
Kevlar composite window. A thin-walled aluminium beam pipe of 16 cm
outer diameter traversing the centre of the spectrometer (and all
the following detectors) allows the undecayed beam particles and the
muon halo from decays of beam pions to continue their path in
vacuum. The spectrometer consists of four drift chambers (DCH):
DCH1, DCH2 located upstream, and DCH3, DCH4 downstream of a dipole
magnet. The magnet has a field integral $\int B_ydz=0.4$~Tm, thus
providing a horizontal transverse momentum kick $\Delta P_x=120~{\rm
MeV}/c$ for charged particles. The DCHs have the shape of a regular
octagon with a transverse size of about 2.8 m and a fiducial area of
about 4.5 m$^2$. Each chamber is composed of eight planes of sense
wires arranged in four pairs of staggered planes oriented
horizontally, vertically, and along each of the two orthogonal
$45^\circ$ directions. The spatial resolution of each DCH is
$\sigma_x=\sigma_y=90\mu$m. The nominal momentum resolution of the
magnetic spectrometer is parameterized as $\sigma_p/p = (1.02 \oplus
0.044\cdot p)\%$ ($p$ expressed in GeV/$c$). The measured resolution
on the reconstructed $3\pi^\pm$ invariant mass varied during the
running period corresponding to the considered data sample in the
range of $(1.65-1.72)$~MeV/$c^2$, depending on DCH performance.

The magnetic spectrometer is followed by a plastic scintillator
hodoscope (HOD) used to produce fast trigger signals and to provide
precise time measurements of charged particles. The hodoscope has a
regular octagonal shape with a transverse size of about 2.4 m. It
consists of a plane of horizontal and a plane of vertical
strip-shaped counters. Each plane consists of 64 counters arranged
in four quadrants. Each quadrant is subdivided into 4 segments
taking part in the trigger logic. Counter widths (lengths) vary from
6.5 cm (121 cm) for central counters to 9.9 cm (60 cm) for
peripheral ones.

The hodoscope is followed by a liquid krypton electromagnetic
calorimeter (LKr), a hadronic calorimeter (HAC) and a muon detector
(MUV), all of which are not used in the present analysis. A detailed
description of the components of the NA48 detector can be found
elsewhere~\cite{ba96}.

The $K^\pm\to\pi^\pm\pi^+\pi^-$ events are triggered with a
two-level system. At the first level (L1), the rate of $\sim 500$
kHz is reduced to $\sim 100$ kHz by requiring coincidences of hits
in the two planes of the HOD in at least two of the 16
non-overlapping segments. The second level (L2) is based on a
hardware system computing coordinates of hits from DCH drift times,
and a farm of asynchronous microprocessors performing fast
reconstruction of tracks and running a selection algorithm, which
requires at least two tracks to originate in the decay volume with
the closest distance of approach less than 5 cm. L1 triggers not
satisfying this condition are examined further and accepted
nevertheless if there is a reconstructed track which is not
kinematically compatible with a $\pi^\pm\pi^0$ decay of a $K^\pm$
having momentum of 60 GeV/$c$ directed along the beam axis. The
resulting trigger rate is about 10 kHz.

%%%%%%%%%%%%%%%%%%%%%%%%%%%%%%%%%%%%%%%%%%%%%%%%%%%%%%%%%%%%%%
\section{Data analysis}

\noindent {\bf Reconstruction and selection} \vspace{2mm}

\noindent Event reconstruction is based entirely on the magnetic
spectrometer information. Tracks are reconstructed from hits in DCHs
using the measured magnetic field map rescaled according to the
recorded value of electric current in the spectrometer analyzing
magnet. Three-track vertices, compatible with a
$K^\pm\to\pi^\pm\pi^+\pi^-$ decay, are reconstructed by
extrapolation of track segments from the upstream part of the
spectrometer back into the decay volume, taking into account the
stray magnetic fields due to the Earth's field and parasitic
magnetization of the vacuum tank, and multiple scattering in the
Kevlar window. The stray field correction is based on a
three-dimensional field map measured in the entire vacuum tank, and
reduces the amplitude of the observed sinusoidal variation of the
reconstructed $3\pi$ invariant mass on the azimuthal orientation of
the odd pion by more than an order of magnitude to a level below
0.05 MeV/$c^2$. The event kinematics is calculated using measured
momenta and track directions extrapolated to the decay vertex.

The principal selection criteria applied to the reconstructed
variables are the following:
\begin{itemize}
\item Total charge of the three pion candidates: $Q=\pm1$;
\item Transverse momentum: $P_T<0.3$~GeV/$c$;
\item Longitudinal vertex position within the decay volume:
$Z_{vtx}>Z_{final~coll.}$;
\item Transverse vertex radius within the beam area: $R_{vtx}<3$~cm;
\item Kaon momentum within the nominal range:
$54~{\rm GeV}/c<|\vec P_K|<66~{\rm GeV}/c$.
\end{itemize}

To improve the resolution on the kinematic variables, and to reduce
the impact of differences between data and MC resolutions, the
events were passed through a kinematic fitting procedure with three
constraints (constraining the initial kaon direction to by along the
$z$ axis, and the $3\pi$ invariant mass to the kaon mass). Events
with the quality of the kinematic fit corresponding to probability
$p<10^{-5}$ were rejected. The fraction of these rejected events
increases as a function of deviation of the reconstructed $3\pi$
mass from the PDG kaon mass $|\Delta M|$; in particular, all the
events with $|\Delta M|>10$~MeV/$c^2$ are outside the signal region,
i.e. rejected by the above condition.

\begin{figure}[tb]
\begin{center}
\resizebox{0.94\textwidth}{!}{\includegraphics{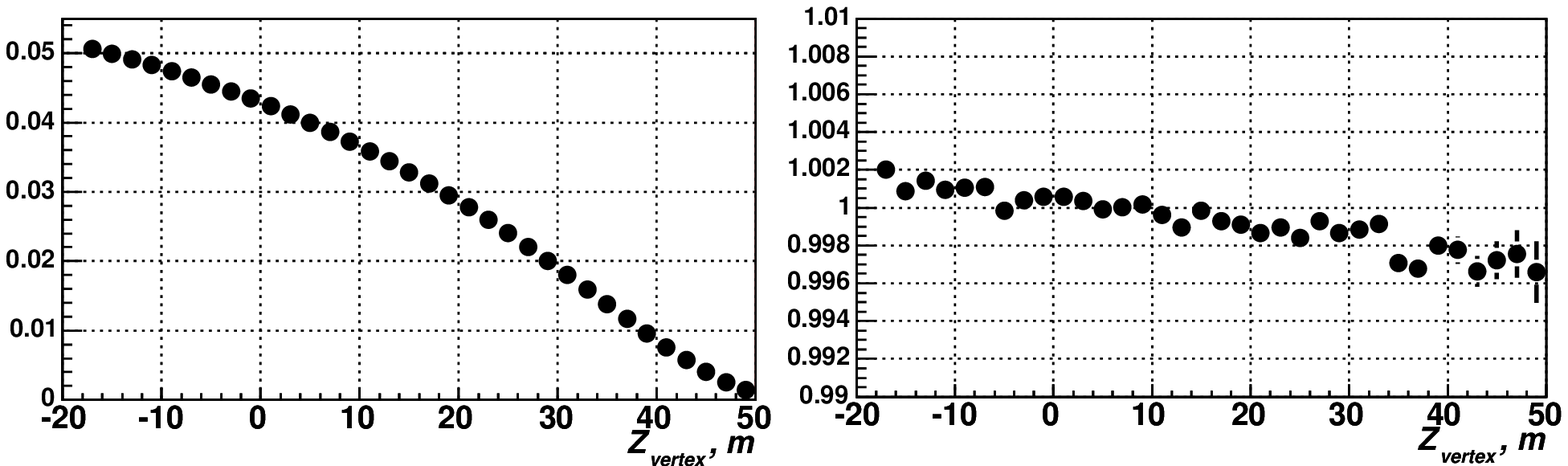}}
\put(-240,100){\Large\bf a}\\
\resizebox{0.94\textwidth}{!}{\includegraphics{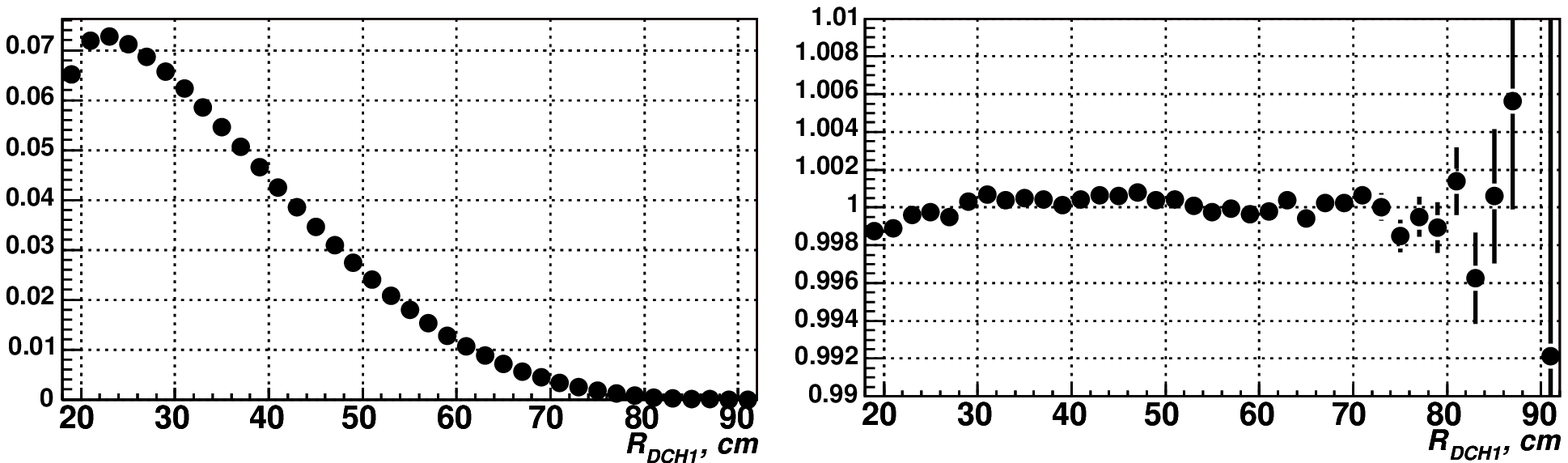}}
\put(-240,100){\Large\bf b}\\
\resizebox{0.94\textwidth}{!}{\includegraphics{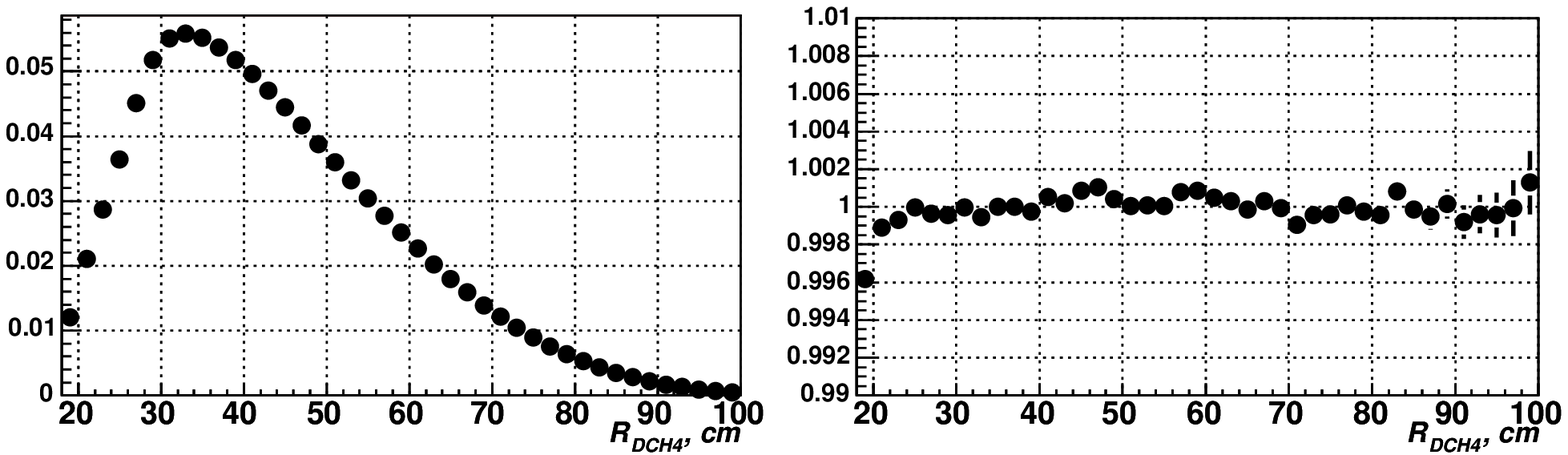}}
\put(-240,100){\Large\bf c}
\end{center}
\vspace{-8mm} \caption{Left column: reconstructed data
distributions, right column: the corresponding ratios of data/MC
distributions of (a) vertex $z$ position, and pion radial position
(for each of the 3 pions) in the planes of (b) DCH1 and (c) DCH4.}
\label{fig:datamc}
\end{figure}

The geometric acceptance for the $K^\pm\to\pi^\pm\pi^+\pi^-$ decays
is mainly determined by the beam pipe traversing the centres of the
DCHs, and the material in the central region of each DCH where
central DCH wires terminate. This material defines a region of high
DCH inefficiency\footnote{The acceptance is not biased by the finite
outer size of the DCHs due to a relatively small $Q$-value of the
$K^\pm\to\pi^\pm\pi^+\pi^-$ decay: $Q=75.0$ MeV.}. This inefficiency
together with beam optics performance and variations influences the
acceptance, and is difficult to reproduce accurately with a MC
simulation. To minimize the effects of this problem, it is required
that the transverse positions of each pion in DCH1 and DCH4 planes
$\vec R_{\pi i}$ ($i=1,4$) satisfy the condition $|\vec R_{\pi
i}-\vec R_{0}|>18$~cm, where $\vec R_{0}$ is the position of the
momentum-weighted average of the three pions' impact points: $\vec
R_{0}=\sum_{i=1}^3 (\vec R_{\pi i}|\vec P_{\pi i}|)/\sum_{i=1}^3
|\vec P_{\pi i}|$ (for DCH4 plane, trajectories of pions are
linearly extrapolated from DCH1 and DCH2 planes). $\vec R_{0}$
corresponds to the transverse position of the line of flight of the
initial kaon. The value of 18 cm was chosen to exclude safely the
inefficient central region taking into account the beam sizes and
variations of their average transverse positions. The described
selection criterion costs about $50\%$ of the statistics; however an
appropriate MC description of the experimental conditions is more
important than the sample size for the present analysis\footnote{In
a different case, the charge asymmetry analysis~\cite{k3pic} is
performed with soft cuts maximizing the selected data sample, and is
based on cancellation, rather than simulation, of the systematic
effects.}. As will be shown below in the discussion leading to the
results presented in Fig.~\ref{fig:datamc}, the MC simulation of the
experimental conditions reproduces the data distributions to a level
of a few parts per mille.

The selection leaves a sample of $4.71\times 10^8$ events, which is
practically background free, as $K^\pm\to\pi^\pm\pi^+\pi^-$ is by
far the dominant decay mode of the charged kaon with more than one
charged particle in the final state. The fact that backgrounds due
to other decays of beam kaons and pions are negligible was also
checked with a MC simulation.

\begin{figure}[tb]
\begin{center}
\begin{tabular}{@{}l@{}r@{}}
{\put(-2,100){
\resizebox{0.47\textwidth}{!}{\includegraphics{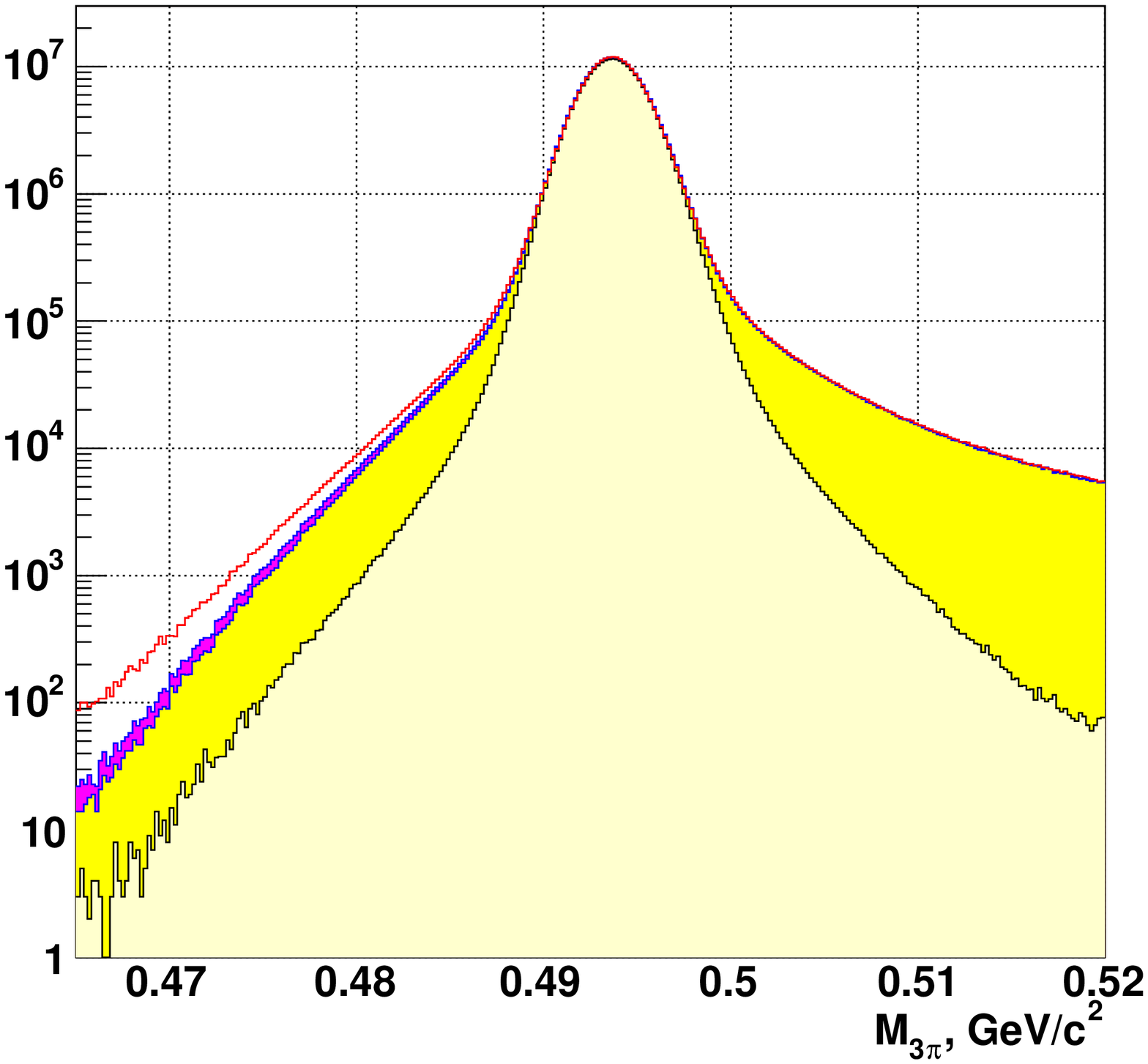}}}
\resizebox{0.47\textwidth}{!}{\includegraphics{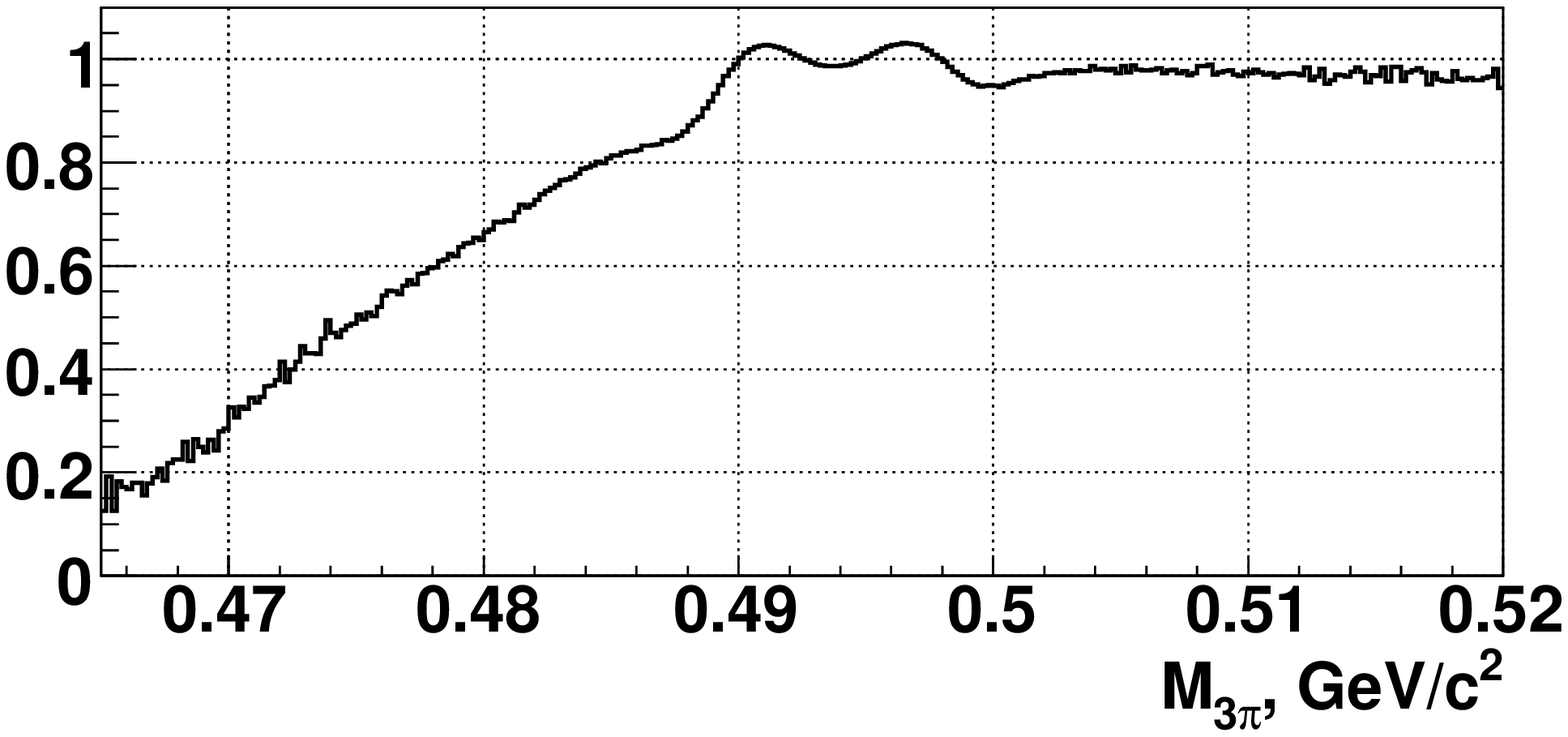}}
\put(-30,148){(1)}\put(-30,184){(2)}\put(-195,190){(3)}
\put(-176,273){\Large\bf a} \put(-176,79){\Large\bf b}
\put(-188,188){\vector(1,-2){8}}} &
\put(7,40){{\resizebox{0.53\textwidth}{!}{\includegraphics{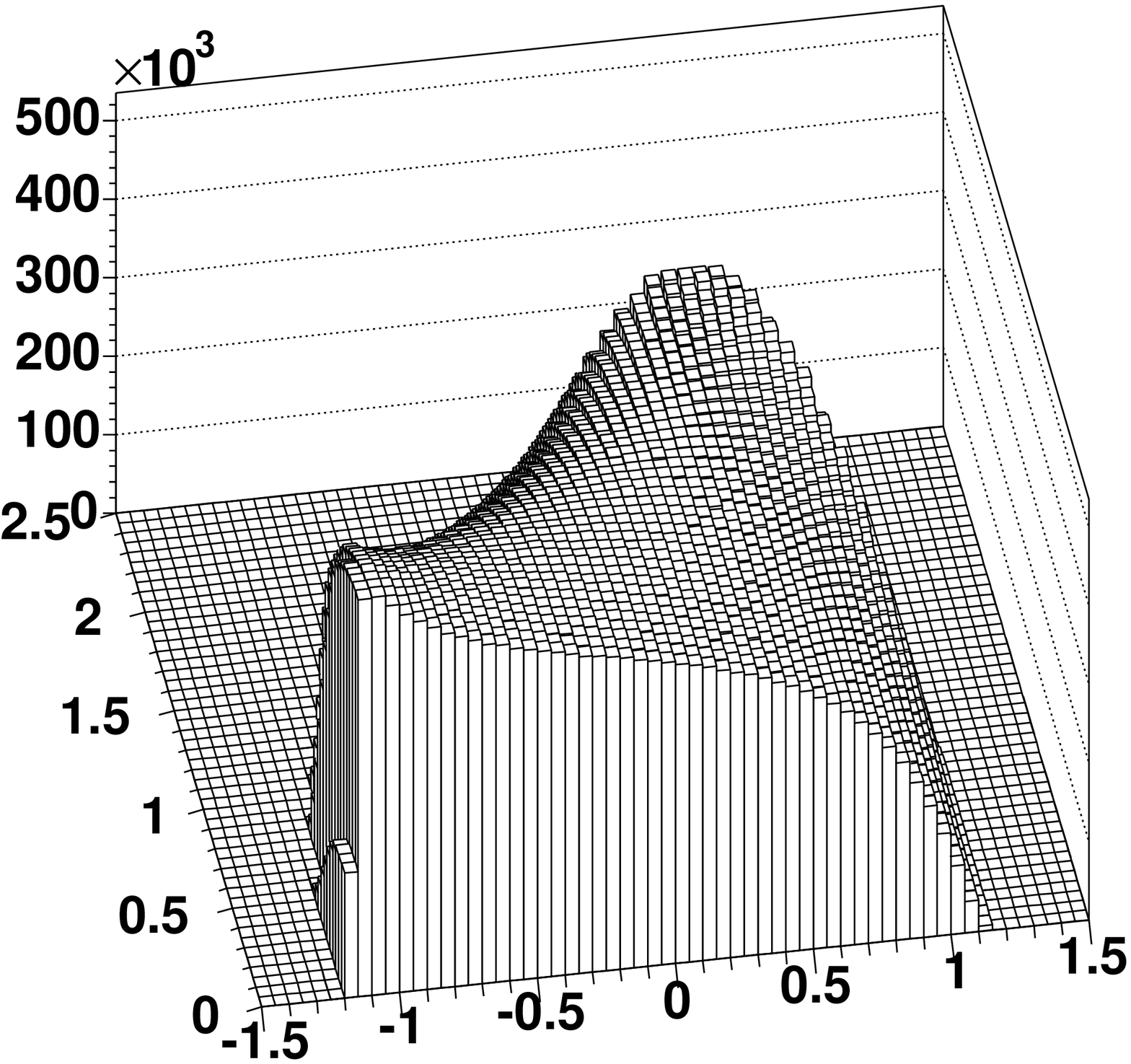}}}
\put(-30,25){$\bf u$}\put(-240,107){$\bf |v|$}
\put(-195,196){\Large\bf c}} {\rule{0.53\textwidth}{0mm}}
\\
\end{tabular}
\end{center}
\vspace{-8mm} \caption{(a) Reconstructed spectrum of $3\pi^\pm$
invariant mass $M(3\pi)$ (upper line) and its presentation in terms
of normalized MC components: (1) events without $\pi\to\mu\nu$ decay
in flight, (2) events with $\pi\to\mu\nu$ decay, (3) IB radiative
$K_{3\pi\gamma}$ events. (b) The ratio of MC/data $M(3\pi)$ spectra
demonstrating the imperfection of resolution description in MC, and
a deficit of MC events at low $M(3\pi)$ outside the signal region.
(c) Reconstructed distribution in the kinematic variables
$F_{data}(u,|v|)$ (after kinematic fit).} \label{fig:kmass}
\end{figure}

The distribution of the reconstructed $3\pi^\pm$ invariant mass of
data events (before the kinematic fitting) and its comparison with
MC are presented in Fig.~\ref{fig:kmass}a. The non-Gaussian tails of
the mass distribution are primarily due to
$\pi^\pm\to\mu^\pm\nu_\mu$ decays in flight, and are well understood
in terms of MC simulation. The ratio of MC to data mass spectra is
presented in Fig.~\ref{fig:kmass}b. It demonstrates the imperfection
of resolution description in MC, and a deficit of MC events in the
low mass region, which however contains a small fraction of the
events, and is mostly outside the signal region. The Dalitz plot
distribution of the selected data events (after the kinematic
fitting) $F_{data}(u,|v|)$ used for the subsequent analysis is
presented in Fig.~\ref{fig:kmass}c. The bin sizes of the Dalitz plot
distributions used in the analysis are $\delta u=\delta v=0.05$.

\vspace{2mm}

\noindent {\bf Correction for trigger inefficiency} \vspace{2mm}

\noindent To simplify the treatment of the trigger inefficiency,
stable trigger performance was the main condition used to select the
sample to be used for the analysis\footnote{As it was already noted,
the size of the data sample is not a limitation for this analysis.}.
Inefficiencies of both L1 and L2 trigger components were directly
measured as functions of ($u,|v|$) using control data samples of
prescaled low bias triggers collected along with the main triggers,
which allowed a correction of the observed ($u,|v|$) distributions,
and propagation of the statistical errors of trigger inefficiencies
into the result.

The L1 trigger condition requiring a coincidence of hits in two of
the 16 non-overlapping HOD segments is loose, as there are three
charged particles in a fully reconstructed event, providing a rather
low (and stable in time) inefficiency of $0.6\times 10^{-3}$ for the
selected event sample. However, the L1 inefficiency depends rather
strongly on the kinematic variables. The primary mechanism
generating such a dependence is the enhancement of inefficiency for
topologies with two pions hitting the same HOD segment; such events
preferably belong to the kinematic regions characterized by small
relative velocity of a certain $\pi\pi$ pair in the kaon rest frame.

The L2 inefficiency, which is due to local DCH inefficiencies
affecting the trigger more strongly than the off-line reconstruction
due to lower redundancy and trigger timing effects, was measured to
be $(0.32\pm0.05)\times 10^{-2}$ (the error indicates the maximum
size of its variation during the data taking period). It did not
exhibit any significant correlation to the kinematic variables, due
to relative complexity of the decision-making algorithm.

\vspace{2mm}

\noindent {\bf Monte Carlo simulation} \vspace{2mm}

\noindent A detailed GEANT-based MC simulation was developed, which
includes full detector geometry and material description, simulation
of stray magnetic fields, DCH local inefficiencies and misalignment,
the beam line (which allows a reproduction of the kaon momentum
spectra and beam profiles), and $K^+/K^-$ relative fluxes. Moreover,
time variations of the above effects during the running period were
simulated. A production of $6.7\times 10^9$
$K^\pm\to\pi^\pm\pi^+\pi^-$ events distributed according to the
matrix element (\ref{slopes}) with PDG values of the slope
parameters~\cite{pdg} was performed to determine the detector
acceptance. A sample of $1.16\times10^9$ MC events (almost 2.5 times
larger than the data sample) passes the selection.

Comparison of data and MC distributions in such significant
variables as longitudinal decay vertex position and illuminations of
DCH1 and DCH4 planes by pions is presented in Fig.~\ref{fig:datamc},
and demonstrates that MC simulation reproduces the data
distributions to a level of a few units of $10^{-3}$. The precision
of the data description can be improved by tighter cuts on pion
radial positions.

\vspace{2mm}

\noindent {\bf Fitting procedure} \vspace{2mm}

\begin{figure}[tb]
\vspace{-7mm}
\begin{center}
\resizebox{0.90\textwidth}{0.32\textheight}{\includegraphics{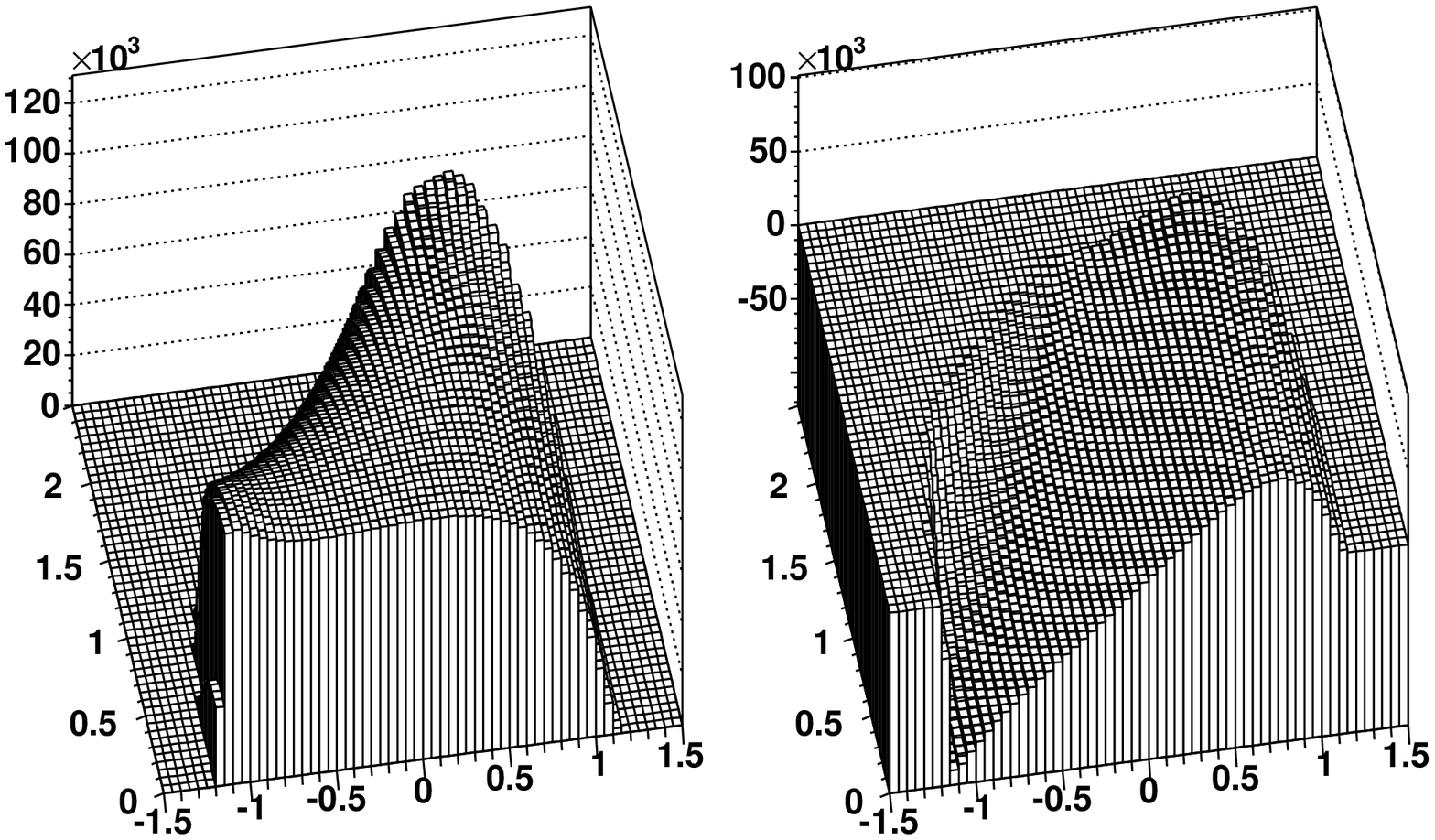}}
\put(-371,180){\Large\bf a} \put(-160,180){\Large\bf b}
\put(-27,18){$\bf u$}\put(-234,18){$\bf u$} \put(-203,105){$\bf
|v|$}\put(-408,103){$\bf |v|$}
\\
\vspace{-4mm}
\resizebox{0.90\textwidth}{0.32\textheight}{\includegraphics{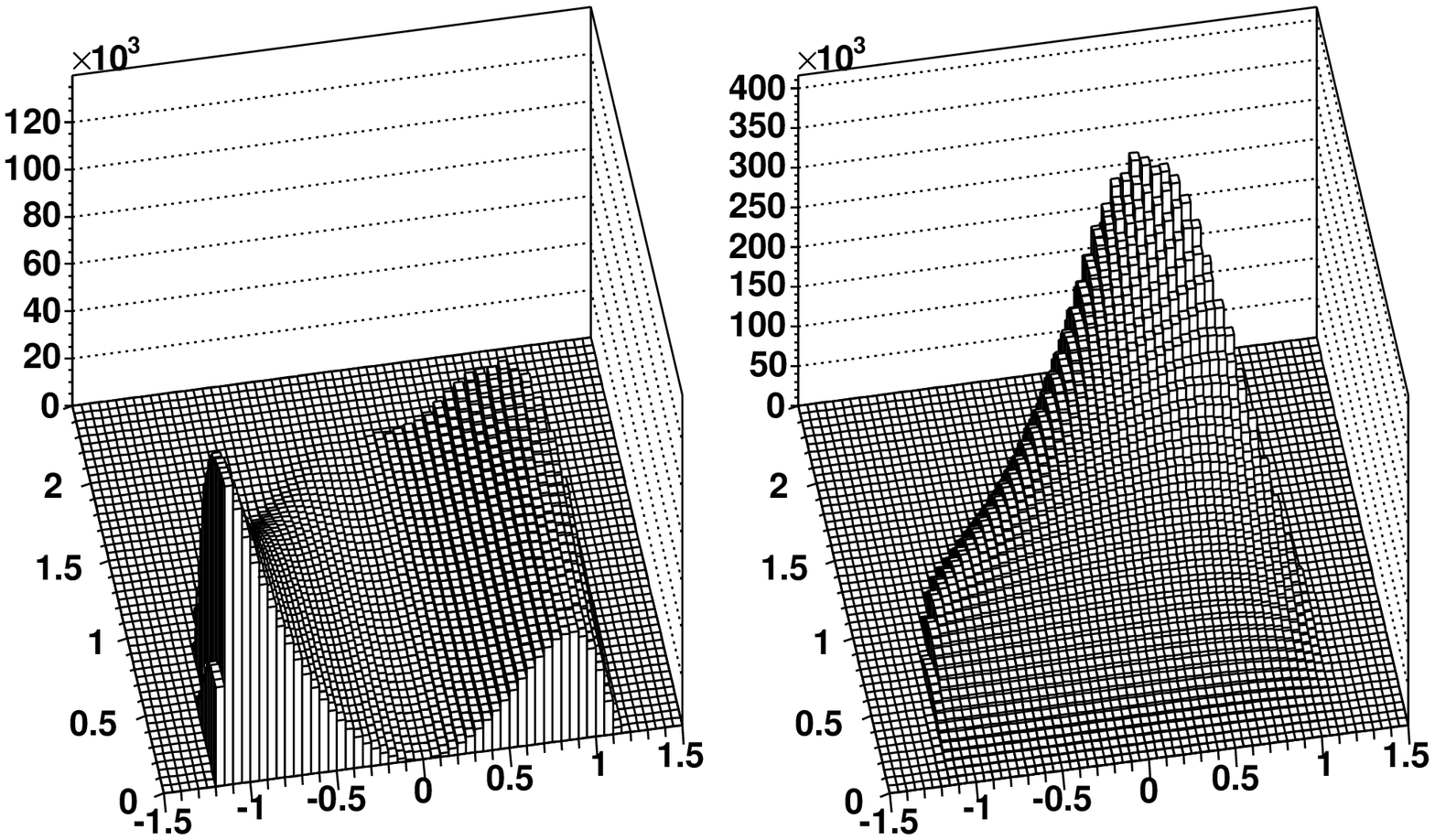}}
\put(-371,180){\Large\bf c} \put(-160,180){\Large\bf d}
\put(-27,18){$\bf u$}\put(-234,18){$\bf u$} \put(-203,103){$\bf
|v|$}\put(-408,103){$\bf |v|$}
\end{center}
\vspace{-11mm} \caption{Reconstructed distributions in the kinematic
variables ($u$,$|v|$) of the four MC components: (a) $F_0(u,|v|)$,
(b) $F_u(u,|v|)$, (c) $F_{u^2}(u,|v|)$, (d) $F_{v^2}(u,|v|)$.}
\label{fig:mc3d}
\end{figure}

\noindent The measurement method is based on fitting the binned
reconstructed data distribution $F_{data}(u,|v|)$ presented in
Fig.~\ref{fig:kmass}c with a sum of four reconstructed MC components
generated according to the four terms in the polynomial
(\ref{slopes}) presented in Fig.~\ref{fig:mc3d}. Let us denote these
reconstructed MC distributions as $F_0(u,|v|)$, $F_u(u,|v|)$,
$F_{u^2}(u,|v|)$, and $F_{v^2}(u,|v|)$. To obtain them, the MC
sample (distributed in kinematic variables according to the PDG
slope parameters) was divided into four subsamples\footnote{The
relative sizes of the four subsamples were subject of optimization
in order to minimize the statistical error of the measurement. The
$F_u(u,|v|)$, $F_{u^2}(u,|v|)$, and $F_{v^2}(u,|v|)$ samples are of
equal sizes, while the $F_0(u,|v|)$ sample is 5 times larger than
each of the former three.}, and events in each subsample were
assigned appropriate weights depending on the generated ($u,|v|$) to
obtain the desired distributions.

The following functional defining the agreement of the shapes of
data and MC distributions is minimized using the MINUIT
package~\cite{minuit} in order to measure the values of the slope
parameters $(g,h,k)$:
\begin{equation}
\label{chi2} \chi^2(g,h,k,N)=\sum_{u,|v|~{\rm
bins}}\frac{(F_{data}(u,|v|)-NF_{MC}(g,h,k,u,|v|))^2} {\delta^2
F_{data}(u,|v|)+N^2\delta^2F_{MC}(g,h,k,u,|v|)}.
\end{equation}
The sum is evaluated over all the bins of reconstructed $(u,|v|)$
distributions with at least 1000 data events, which eliminates the
necessity to include non-Gaussian behaviour of errors. Here $\delta
F_{data}(u,|v|)$ is an uncertainty of the number of data events in a
given bin (composed of a statistical part and a trigger efficiency
part added in quadrature), $F_{MC}(g,h,k,u,|v|)$ is a MC population
of a bin for given values of $(g,h,k)$:
\begin{eqnarray}
F_{MC}(g,h,k,u,|v|) &=& F_0(u,|v|)/I_0 + gF_u(u,|v|)/I_{u} +\\
&&hF_{u^2}(u,|v|)/I_{u^2} + kF_{v^2}(u,|v|)/I_{v^2},\nonumber
\end{eqnarray}
and $\delta F_{MC}(g,h,k,u,|v|)$ is its statistical error:
\begin{eqnarray}
\delta^2F_{MC}(g,h,k,u,|v|) &=& \delta^2F_0(u,|v|)/I_0^2 +
g^2\delta^2
F_u(u,|v|)/I_{u}^2 +\\
&&h^2\delta^2F_{u^2}(u,|v|)/I_{u^2}^2 + k^2\delta^2
F_{v^2}(u,|v|)/I_{v^2}^2.\nonumber
\end{eqnarray}
Here $I_0$, $I_u$, $I_{u^2}$ and $I_{v^2}$ are the normalization
constants computed taking into account the numbers of generated
events in each of the four MC subsamples, and the integrals of the
four terms in (\ref{slopes}) over the Dalitz plot. The free
parameters of the functional (\ref{chi2}) are the slope parameters
$(g,h,k)$ and an overall MC normalization parameter $N$.

The minimization yields $\chi^2/{\rm NDF}=1669/1585$, corresponding
to a satisfactory probability of 7.0\%. The results of the fit and
the trigger corrections are presented in Table~\ref{tab:fit}. The
non-zero values of the corrections arise mostly from the L1 trigger
inefficiency dependence on kinematic variables, while their
statistical errors receive contributions of similar sizes from L1
and L2 trigger efficiency uncertainties.

\begin{table}[tb]
\begin{center}
\begin{tabular}{c|rcl|cc|rcl}
\hline Parameter
&\multicolumn{3}{c|}{Value}&\multicolumn{2}{c|}{Uncertainties}&
\multicolumn{3}{c}{Trigger correction}
\\
\cline{5-6} &&&& statistical & MC stat.&&\\
\hline
$g\times10^2$&$-21.134$&$\pm$&$0.013$&0.009&0.008&$-0.008$&$\pm$&0.005\\
$h\times10^2$&$  1.848$&$\pm$&$0.022$&0.015&0.013&$ 0.116$&$\pm$&0.009\\
$k\times10^2$&$ -0.463$&$\pm$&$0.007$&0.005&0.004&$ 0.033$&$\pm$&0.003\\
\hline
\end{tabular}
\end{center}
\vspace{-5mm} \caption{Fit results with statistical and MC
statistical uncertainties, trigger corrections and their statistical
uncertainties. The trigger corrections are included into the
resulting values.} \label{tab:fit}
\end{table}

Keeping only the linear term $gu$ in (\ref{slopes}) yields a fit of
unacceptable quality: $\chi^2/{\rm NDF}=13683/1583$. Including the
terms proportional to $u^3$ and $uv^2$ (the only cubic terms allowed
by Bose symmetry) yields values for the corresponding cubic slope
parameters compatible with zero.

\vspace{2mm}

\noindent {\bf Stability checks} \vspace{2mm}

\begin{figure}[tb]
\begin{center}
\resizebox{0.49\textwidth}{!}{\includegraphics{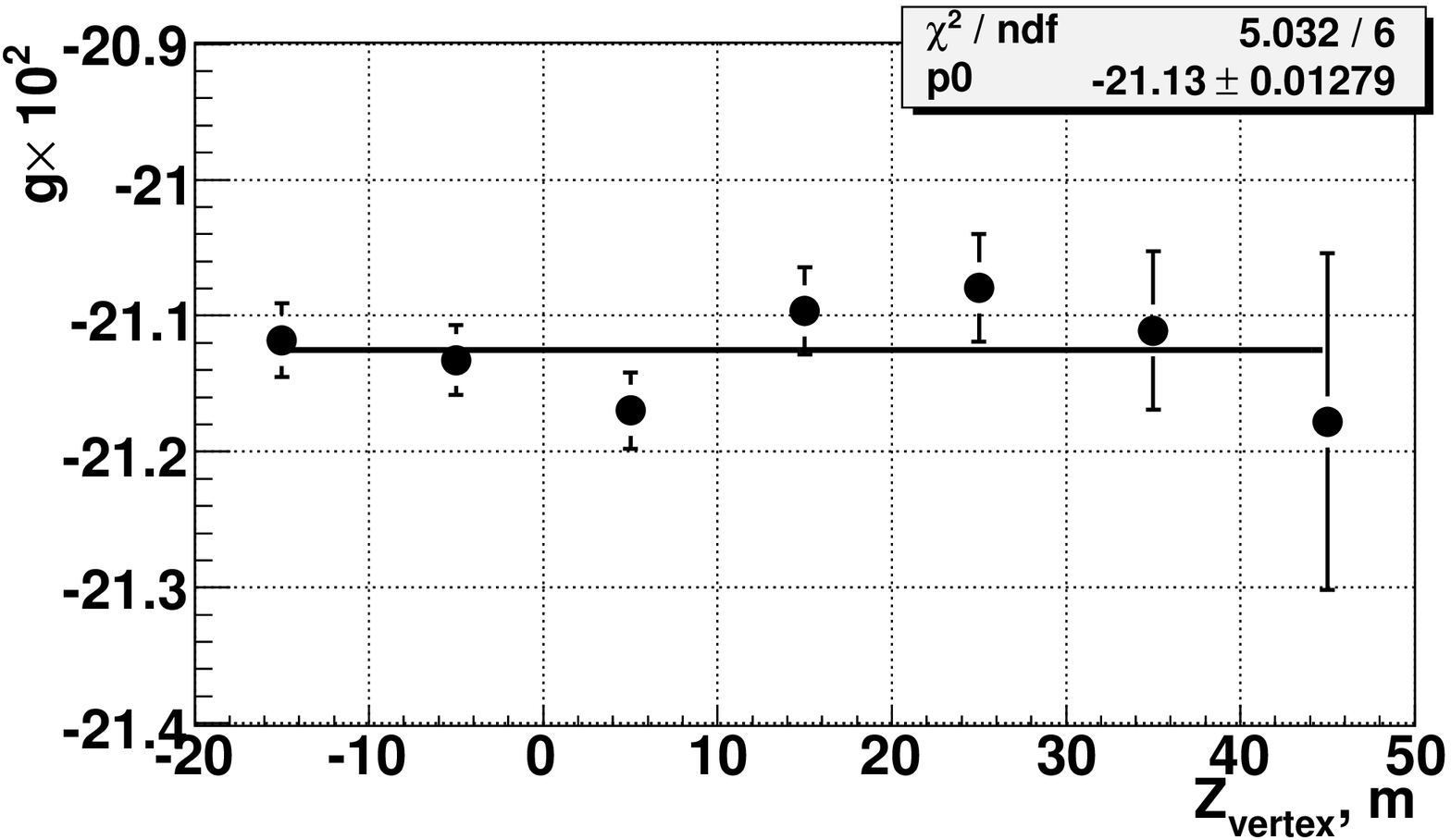}}
\resizebox{0.49\textwidth}{!}{\includegraphics{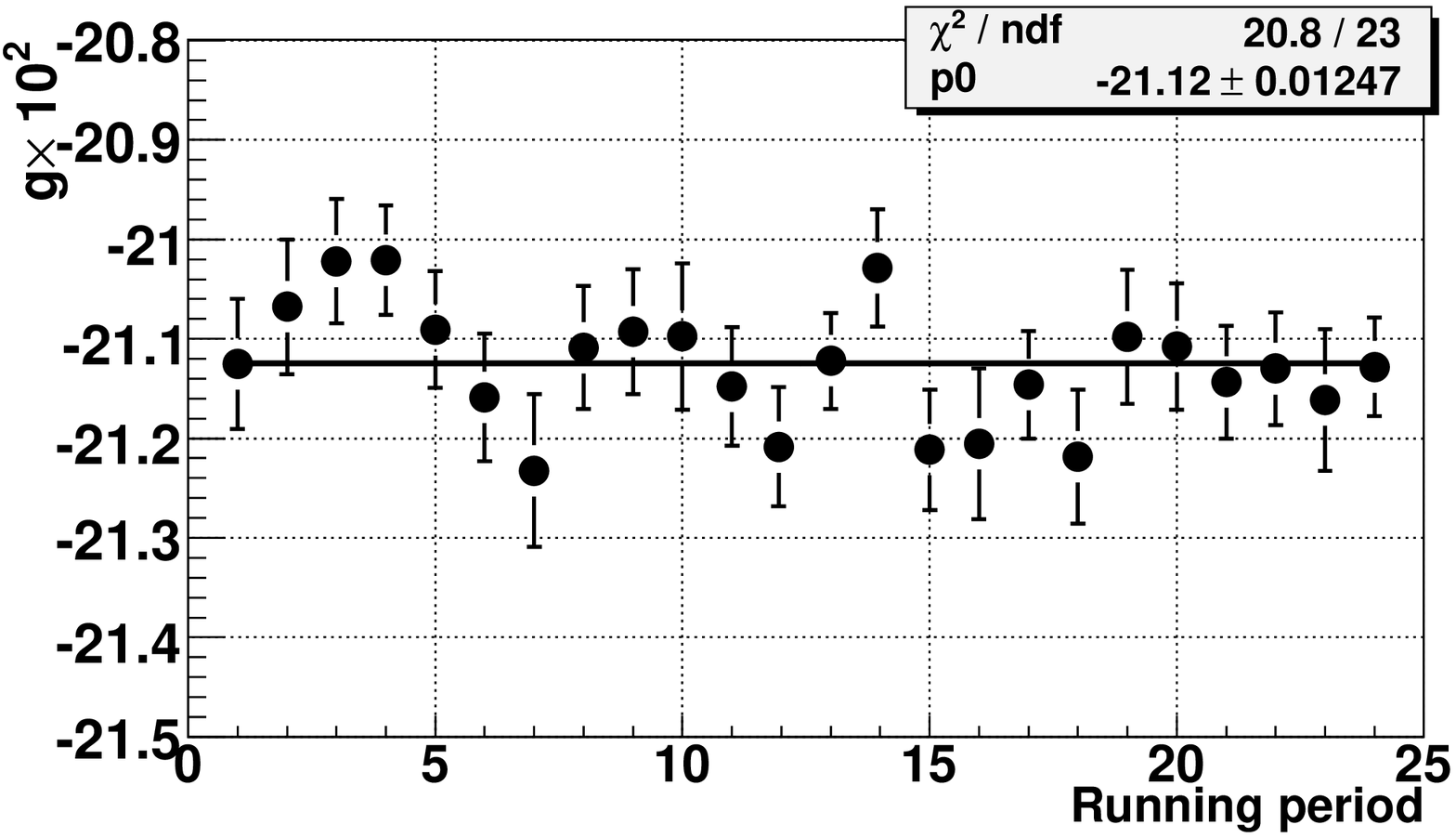}}\\
\resizebox{0.49\textwidth}{!}{\includegraphics{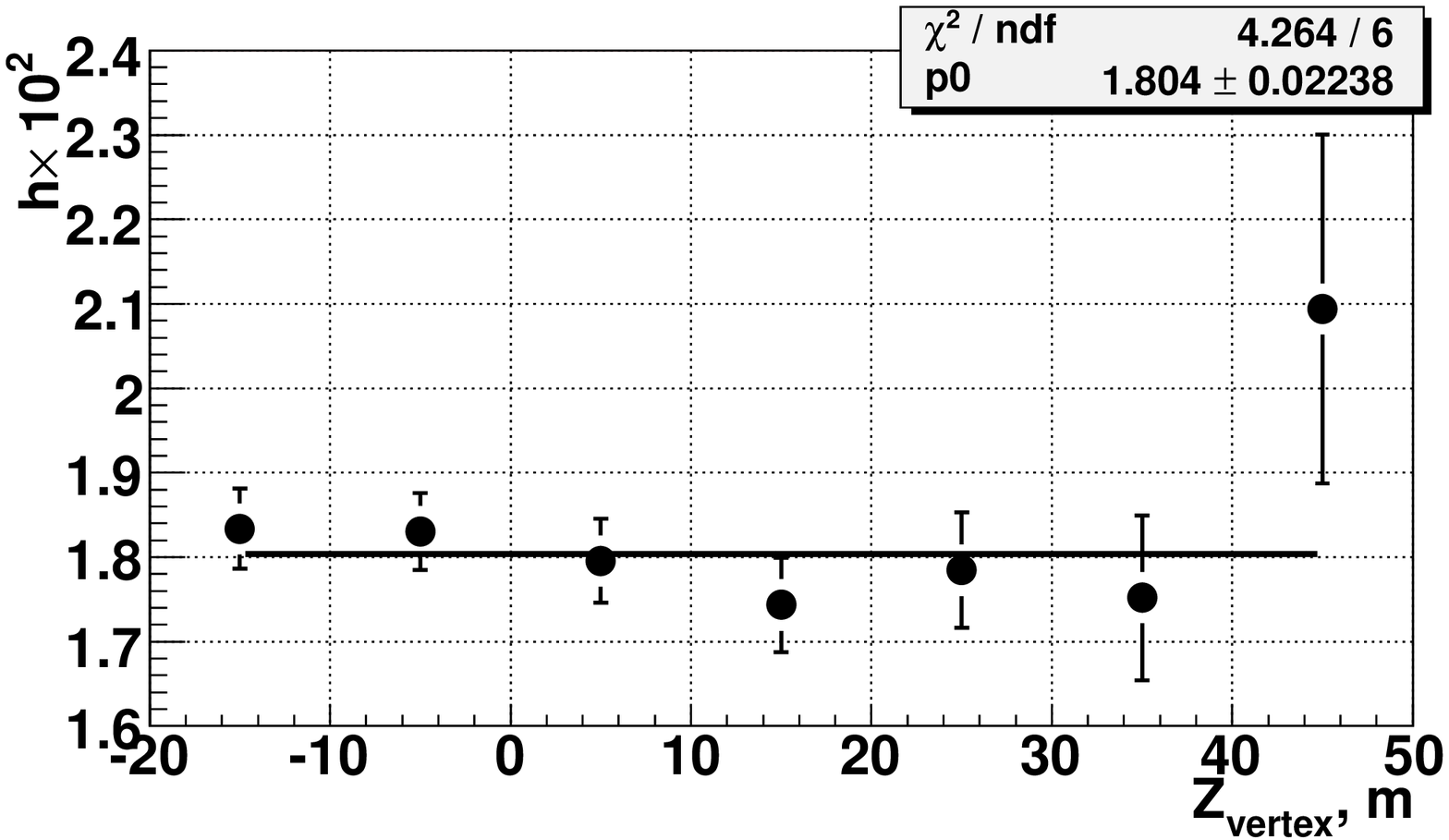}}
\resizebox{0.49\textwidth}{!}{\includegraphics{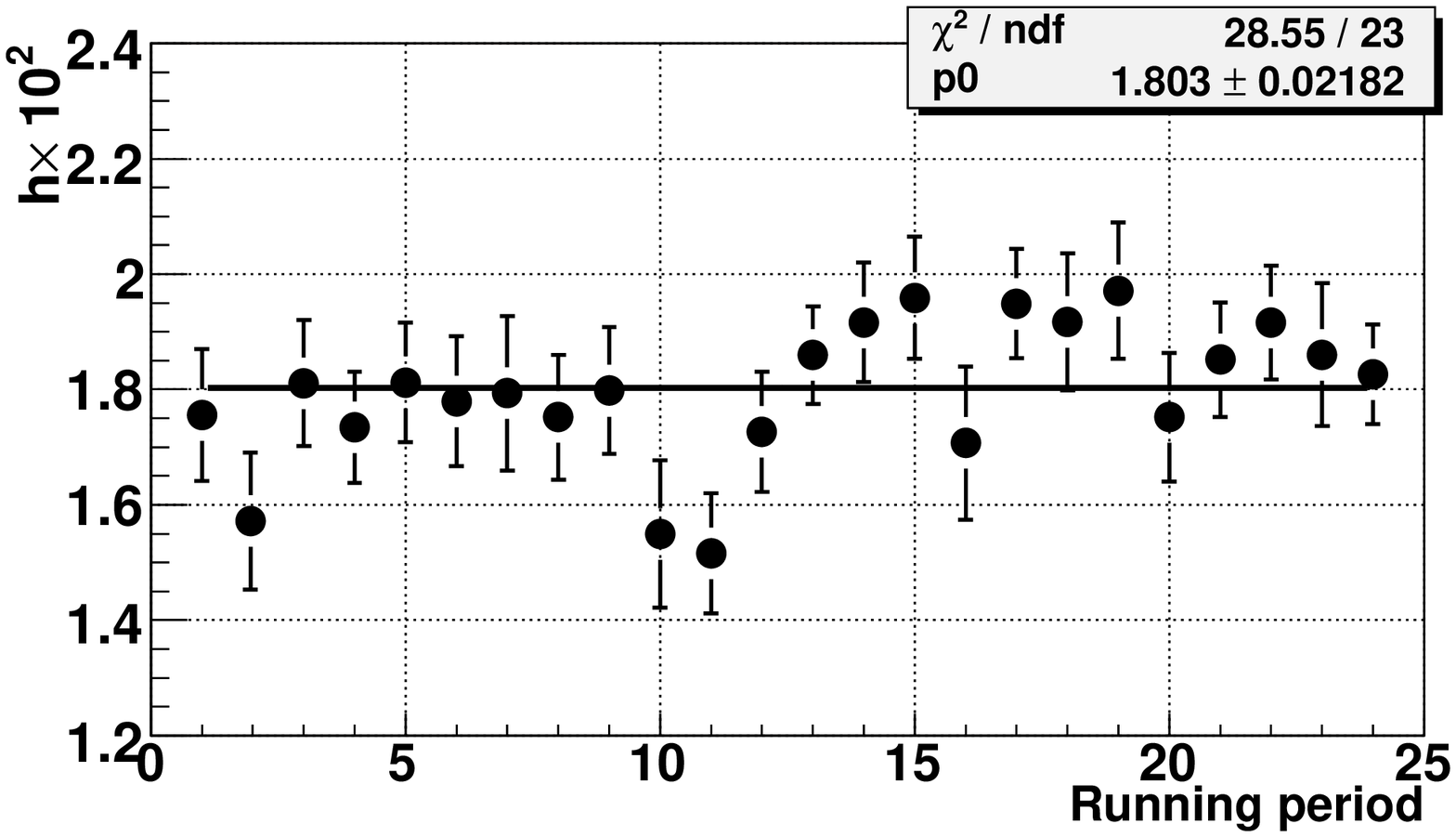}}\\
\resizebox{0.49\textwidth}{!}{\includegraphics{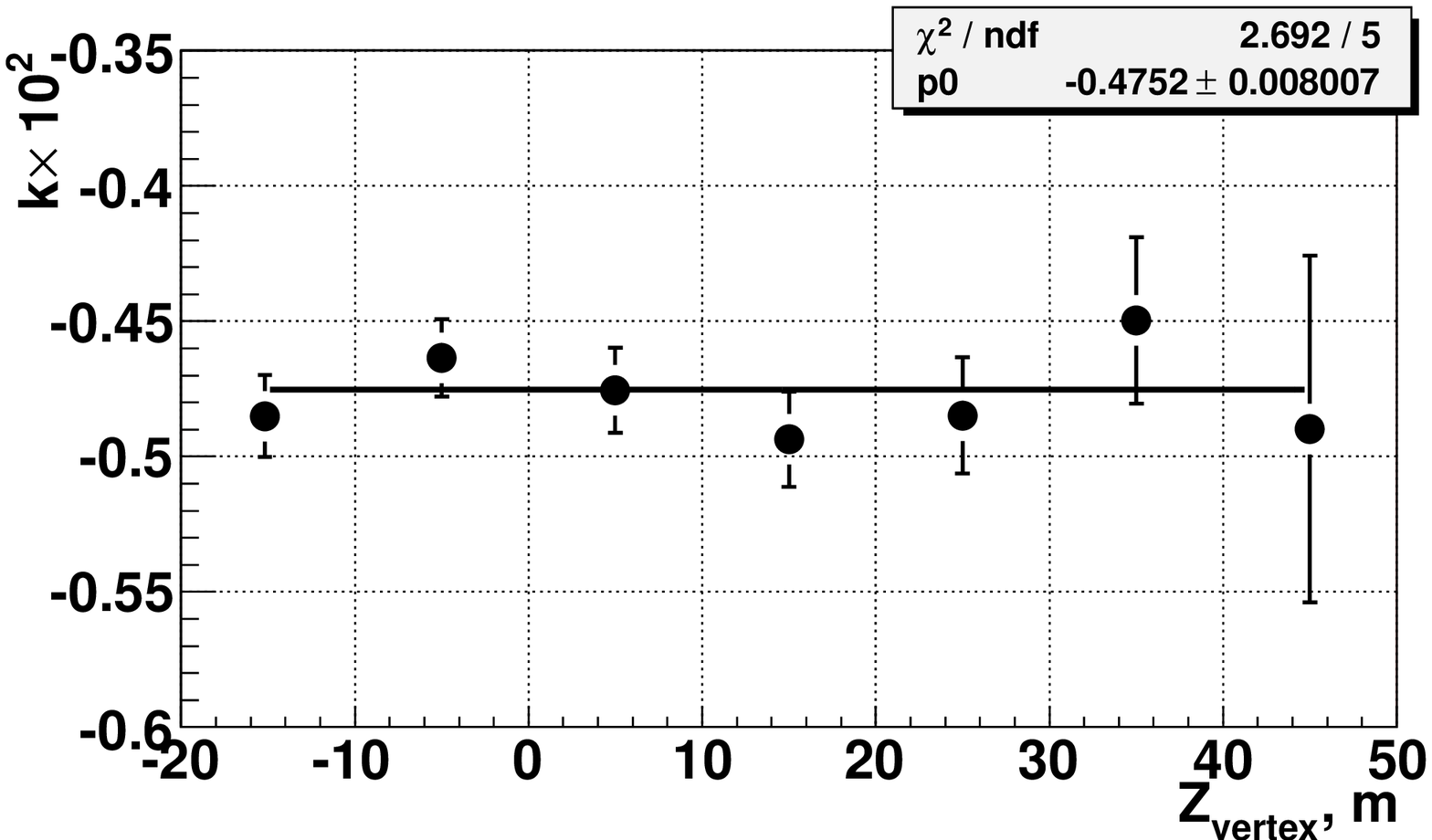}}
\resizebox{0.49\textwidth}{!}{\includegraphics{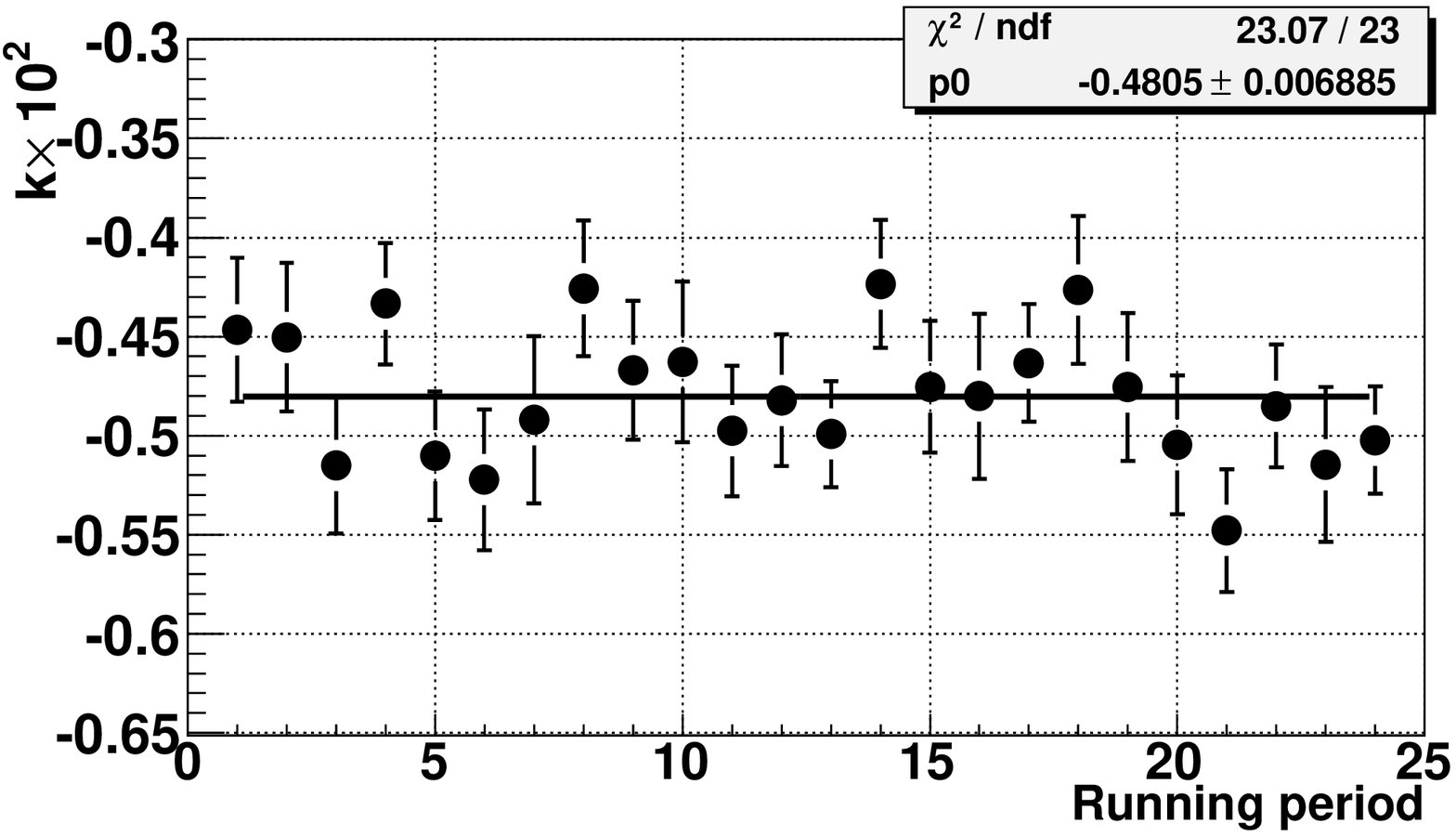}}\\
\end{center}
\vspace{-8mm} \caption{Slope measurements in bins of $Z_{vtx}$ (left
column) and in running periods (right column). Systematic errors are
not shown.} \label{fig:stab}
\end{figure}

\noindent Stability of the results with respect to variations of the
selection conditions on vertex fit quality, $P_T$, $R_{vtx}$, $|\vec
P_K|$ and $\vec R_{\pi i}$, and binning variations was checked.
Stability with respect to exclusion of ($u$, $|v|$) bins with large
deviations of the Coulomb factor (\ref{coulomb}) from unity and with
respect to kaon sign\footnote{Combined $K^+$ and $K^-$ sample is
used to obtain the result. Stability of the slope parameters with
respect to kaon sign is a consequence of the experimental
fact~\cite{k3pic} that CP invariance holds at the discussed level of
precision.} was checked as well. No statistically significant
dependencies were found. Stability with respect to various ways of
binning the data was checked; comparison of slope measurements in
bins of reconstructed longitudinal coordinate of decay vertex
$Z_{vtx}$ (since the acceptance depends strongly on this variable),
and in data taking periods are shown in Fig.~\ref{fig:stab} as the
most significant examples.

\vspace{2mm}

\noindent {\bf Systematic uncertainties} \vspace{2mm}

\noindent The Coulomb factor (\ref{coulomb}) used in description of the
event density contains a pole $C(u,v)\to\infty$ corresponding to a
pair of opposite sign pions having zero relative velocity. The
implemented Monte Carlo simulation involves a certain approximation
to $C(u,v)$ in the pole region. In view of that,
sensitivity of the result to
treatment of the pole was studied, in particular, by using an
alternative fitting method involving projections of the distributions
in each of kinematic variables. The assigned conservative systematic
uncertainties due to the fitting procedure are listed in
Table~\ref{tab:syst}. They are expected to be diminished to
a negligible level in a future analysis of the full statistics.

\begin{table}[tb]
\begin{center}
\begin{tabular}{l|rrr}
\hline Effect&~~~$\delta g\times10^2$&~~~$\delta h\times10^2$&~~~$\delta k\times10^2$\\
\hline
Fitting procedure              &0.009~~~&0.007~~~&0.006~~~\\
Pion momentum resolution       &0.004~~~&0.031~~~&0.009~~~\\
Spectrometer magnetic field    &0.002~~~&0.008~~~&0.004~~~\\
Spectrometer misalignment      &0.002~~~&0.002~~~&0.001~~~\\
Stray magnetic field           &0.001~~~&0.002~~~&0.001~~~\\
\hline
Total                          &0.010~~~&0.033~~~&0.012~~~\\
\hline
\end{tabular}
\end{center}
\vspace{-5mm} \caption{A summary of the systematic uncertainties.}
\label{tab:syst}
\end{table}

An important source of systematic uncertainty is the imperfect
description of the resolution in pion momentum, which can be
observed in Fig.~\ref{fig:kmass}b as a slight disagreement of the
shapes of the reconstructed $M(3\pi)$ spectra for data and MC in the
signal region\footnote{As was discussed above, a large disagreement
at low $M(3\pi)$ values is outside the signal region.}. To evaluate
the corresponding effect, two different plausible ways of
introducing smearing of the MC resolution were used: either by
increasing the smearing of DCH space points from 90$\mu$m to
100$\mu$m, or by adding an extra $0.09\%X_0$ layer of matter in the
position of the Kevlar window (the former correction is more
realistic). The sizes of the added perturbations are such as to
correct for the ``double bump'' shape of the ratio in the signal
region. Both methods lead to similar systematic uncertainties on the
slope parameters attributed to description of resolution in pion
momentum. These uncertainties are listed in Table~\ref{tab:syst}.

Effects due to imperfect knowledge of the magnetic field in the
spectrometer magnet were evaluated. The variation of the magnet
current can be monitored with a relative precision of $5\times
10^{-4}$. Smaller variations are continuously controlled with a
precision of $\sim 10^{-5}$ by the deviation of the measured
charge-averaged kaon mass from the nominal PDG value. A
time-dependent correction is introduced by scaling the reconstructed
pion momenta, decreasing the effect of overall field scale to a
negligible level. To account for possible differences between the
shape of the field map used for simulation and the true field,
variations of the MC field map were artificially introduced,
consistently with the known precision of field measurement of $\sim
10^{-3}$. The corresponding uncertainties are listed in
Table~\ref{tab:syst}.

The transverse positions of DCHs and individual wires were
controlled and realigned at the level of reconstruction software
every 2--4 weeks of data taking using data collected in special runs
in which muon tracks were recorded with no magnetic field in the
spectrometer. This allows an alignment precision of $\sim30~\mu$m to
be reached. However, time variations of DCH alignment on a shorter
time scale can bias the measurement. These variations were measured
by the difference between the average reconstructed $3\pi$ invariant
masses for $K^+$ and $K^-$ decays, and taken into account. The
precision with which these effects are simulated leads to systematic
uncertainties presented in Table~\ref{tab:syst}.

The effects due to limited precision of measurement of the stray
magnetic field in the decay volume were estimated by variation of
the stray field map used for decay vertex reconstruction; the
corresponding systematic effects are presented in
Table~\ref{tab:syst}.

The kaon momentum spectra were carefully simulated, and the related
residual uncertainties were found to be negligible.
Possible differences between
data and MC transverse scales were found to have a negligible
influence on the result.

The total systematic errors were obtained by summing the above
contributions in quadrature, and are presented in
Table~\ref{tab:syst}.

%%%%%%%%%%%%%%%%%%%%%%%%%%%%%%%%%%%%%%%%%%%%%%%%%%%%%%%%%%%
\section*{Conclusions}

The Dalitz plot slope parameters of the $K^\pm\to\pi^\pm\pi^+\pi^-$
decays measured with a fraction of NA48/2 data sample ignoring
radiative effects (apart from the Coulomb factor) and strong
rescattering effects, are:
$$
g=(-21.134\pm0.017)\%,~~ h=(1.848\pm0.040)\%,~~k=(-0.463\pm0.014)\%.
$$
These values are in agreement with the world averages\footnote{The
PDG averages results separately for $K^+$ and $K^-$~\cite{pdg};
averaging the PDG data between $K^+$ and $K^-$ decays should take
into account correlated systematic uncertainties of the $K^+$ and
$K^-$ measurements by the same experiment~\cite{fo72}.}, and have an
order of magnitude smaller uncertainties. This is the first
measurement of a non-zero value of the quadratic slope parameter
$h$. The compatibility of the measured distribution with the PDG
polynomial parameterization~\cite{pdg} appears still to be
acceptable at an improved level of precision; no significant higher
order slope parameters were found.

The measurement of the slope parameters is in agreement with a full
next-to-leading order computation~\cite{ga03}:
$$
g=(-22.0\pm2.0)\%,~~ h=(1.2\pm0.5)\%,~~k=(-0.54\pm0.15)\%.
$$

The whole NA48/2 sample suitable for $K^\pm\to3\pi^\pm$ Dalitz plot
distribution analysis contains at least three times more data; a
more elaborate analysis is foreseen when the corresponding
theoretical framework is available.

%%%%%%%%%%%%%%%%%%%%%%%%%%%%%%%%%%%%%%%%%%%%%%%%%%%%%%%%%%%
\section*{Acknowledgements}

It is a pleasure to thank the technical staff of the participating
laboratories, universities and affiliated computing centres for
their efforts in the construction of the NA48 apparatus, in
operation of the experiment, and in data processing.

%\end{linenumbers}

%

%
\end{document}